# Altered structural connectivity of the left visual thalamus in developmental dyslexia

Christa Müller-Axt[1,*], Alfred Anwander[2], Katharina von Kriegstein[1,3,4]

[1] Research Group Neural Mechanisms of Human Communication, Max Planck Institute for Human Cognitive and Brain Sciences, Leipzig, Germany

[2] Department of Neuropsychology, Max Planck Institute for Human Cognitive and Brain Sciences, Leipzig, Germany

[3] Department of Psychology, Humboldt University of Berlin, Berlin, Germany

[4] Faculty of Psychology, Technical University of Dresden, Dresden, Germany

* Corresponding author:

Christa Müller-Axt

Max Planck Institute for Human Cognitive and Brain Sciences

Research Group Neural Mechanisms of Human Communication

Stephanstr. 1a, 04103 Leipzig, Germany

Phone: +49 341 9940-2306

Email: muelleraxt@cbs.mpg.de

http://www.cbs.mpg.de/employees/muelleraxt

*Preprint submitted to Currrent Biology.*  *Final draft: 09/2017*

Altered Structural Connectivity of the Left Visual Thalamus in Developmental Dyslexia

## Summary

**Developmental dyslexia is a highly prevalent reading disorder affecting about 5-10% of children [1]. It is characterized by slow and/or inaccurate word recognition skills as well as by poor spelling and decoding abilities [2]. Partly due to technical challenges with investigating subcortical sensory structures, current research on dyslexia in humans by-and-large focuses on the cerebral cortex [3-7]. These studies found that dyslexia is typically associated with functional and structural alterations of a distributed left-hemispheric cerebral cortex network [e.g., 8, 9]. However, findings from animal models and post-mortem studies in humans suggest that dyslexia might also be associated with structural alterations in subcortical sensory pathways [10-14, reviewed in ref. 7]. Whether these alterations also exist in dyslexia in-vivo and how they relate to dyslexia symptoms is currently unknown. Here we used ultra-high resolution structural magnetic resonance imaging (MRI), diffusion MRI and probabilistic tractography to investigate the structural connections of the visual sensory pathway in dyslexia in-vivo. We discovered that individuals with dyslexia have reduced structural connections in the direct pathway between the left visual thalamus (LGN) and left middle temporal area V5/MT, but not between the left LGN and left primary visual cortex. In addition, left V5/MT-LGN connectivity strength correlated with rapid naming abilities – a key deficit in dyslexia [15]. These findings provide the first evidence of specific structural alterations in the connections between the sensory thalamus and cortex in developmental dyslexia. The results challenge current standard models and provide novel evidence for the importance of cortico-thalamic interactions in explaining dyslexia.**

Altered Structural Connectivity of the Left Visual Thalamus in Developmental Dyslexia## Results and Discussion

It is a long-standing, but untested, hypothesis that histological alterations of the visual sensory thalamus (i.e., the lateral geniculate nucleus, LGN), as found in dyslexics post-mortem, would be associated with alterations in the structural connections of the visual pathway [10]. However, to date there is no study that has yet examined the structural connections of the LGN in individuals with dyslexia. The reason is that the LGN's small size and deep position within the brain make it difficult to spatially map the LGN using noninvasive imaging techniques. Here we overcame this challenge by using ultra-high resolution 7 Tesla (7T) MRI quantitative $T_1$ maps to individually delineate the LGN in a group of dyslexic adults (n=12, Table S1) and matched control participants (n=12, Table S1) (Figure 1A and 1B). We particularly focused on the LGN in the left hemisphere given evidence that dyslexia is associated with functional and structural alterations particularly of left-hemispheric regions [16, 17, reviewed in ref. 3]. The LGN segmentation procedure resulted in an average left LGN volume of 119 ± 22 mm$^3$ in controls and 114 ± 19 mm$^3$ in dyslexics, which is in good agreement with the average volume of 115 mm$^3$ that has been previously measured for the left LGN in post-mortem human specimens [18].

We tested the hypothesis that developmental dyslexia is associated with reduced structural connections between cerebral cortex areas and the LGN [10]. We targeted two white matter pathways: the structural white matter connections (i) between the LGN and primary visual cortex (V1), and (ii) between the LGN and middle temporal area V5/MT. The choice for targeting the LGN-V1 pathway was motivated by the fact that these connections constitute the primary cortico-subcortical fiber pathway of the visual system, as the majority of LGN neurons are connected with V1 [19, 20]. The choice for targeting the LGN-V5/MT pathway was motivated by two reasons: First, area V5/MT has direct V1-bypassing connections with the LGN [21-25]. Second, area V5/MT has frequently been implicated in the context of dorsal visual stream dysfunction in developmental dyslexia [reviewed in ref. 26].

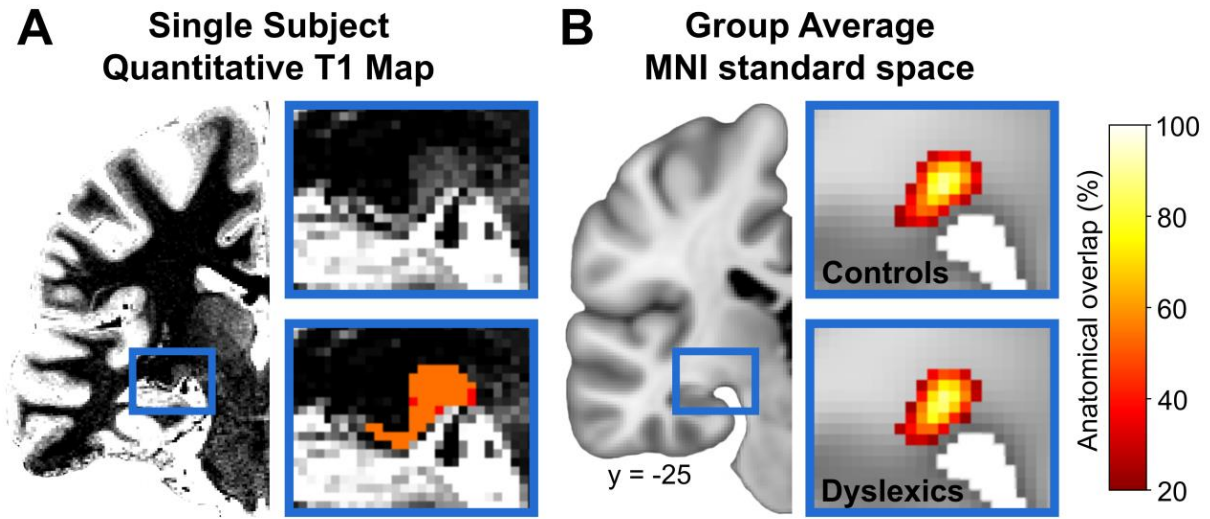

**Figure 1.** Manual segmentations of the left LGN. **A.** Left: Coronal view of the left hemisphere on the quantitative $T_1$ map of a representative single subject. The blue rectangle marks the region, which is shown at higher magnification in the two accompanying inserts. Top right insert: Enlarged view of the left LGN. Bottom right insert: Left LGN overlaid with a conjoined mask created from the manually segmented LGN masks by two independent raters. Orange color depicts voxels that were included by both raters. Red color depicts voxels that were only included by one of the two raters. Segmentations of both raters were merged for each participant, such that only those voxels that were segmented by both raters comprised the final LGN masks used for probabilistic tractography (orange color). The inter-rater reliability of these segmentations was high (mean dice coefficient: 0.86; see Methods). **B.** Left: Coronal view of the left hemisphere on the Montreal Neurological Institute (MNI) standard brain. The blue rectangle marks the region, which is shown at higher magnification in the two accompanying inserts. Top right insert: Group average map of the LGN masks in control participants (n=12) in MNI standard space. Bottom right insert: Group average map of the LGN masks in dyslexic participants (n=12) in MNI standard space. Group average maps were thresholded to show a minimum of 20% anatomical overlap across participants, and showed a high consistency in anatomical LGN location (see Methods). Contrary to a previous report [55], there was no significance difference in LGN volume between the two groups [$t(22)$ =.56, $P$ = .58].

Altered Structural Connectivity of the Left Visual Thalamus in Developmental Dyslexia

We derived the left V1 and V5/MT masks from a volume-based probabilistic atlas (see Methods). Both the V1-LGN (Figure 2A) and the V5/MT-LGN (Figure 2B) connection could be reliably reconstructed by probabilistic tractography in all participants (N=24; see Methods). The direct V5/MT-LGN connection was clearly separate from the V1-LGN connection and was consistently located dorsal to the V1-LGN connection in both groups (Figure 3).

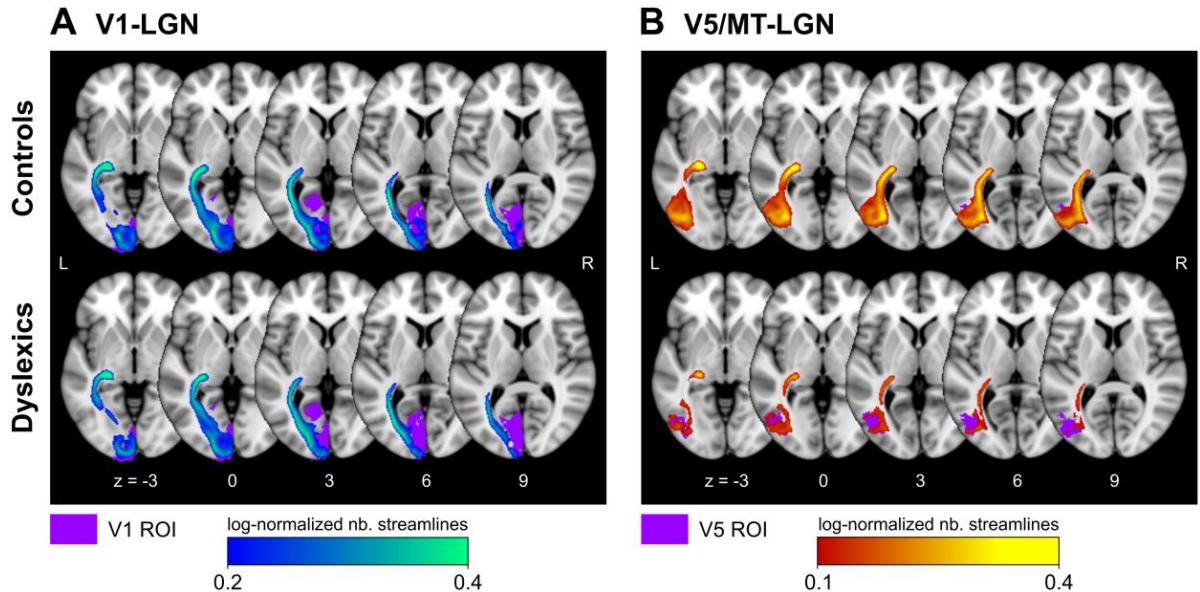

**Figure 2.** Two-dimensional representation of the left V1-LGN and V5/MT-LGN group averaged tracts in controls (n=12) and dyslexics (n=12) in MNI standard space. The group averaged tracts represent the mean probability and strength of a connection computed by the number of reconstructed streamlines per voxel (see Methods). **A.** Left V1-LGN connection in controls and dyslexics. Cortical seed area left V1 is indicated in purple. For visualization purposes, the group averaged tracts were thresholded to show voxels with an average log-normalized number of streamlines per voxel of at least 0.2. **B.** Left V5/MT-LGN connection in controls and dyslexics. Cortical seed area left V5/MT is indicated in purple. Note that cortical seed area V5/MT is barely visible in control participants due to the large extent of resolved V5/MT-LGN streamlines in this group. For visualization purposes, the group averaged tracts were thresholded to show voxels with an average log-normalized number of streamlines per voxel of at least 0.1. Numbers below images indicate z coordinates in MNI standard space.

Altered Structural Connectivity of the Left Visual Thalamus in Developmental Dyslexia

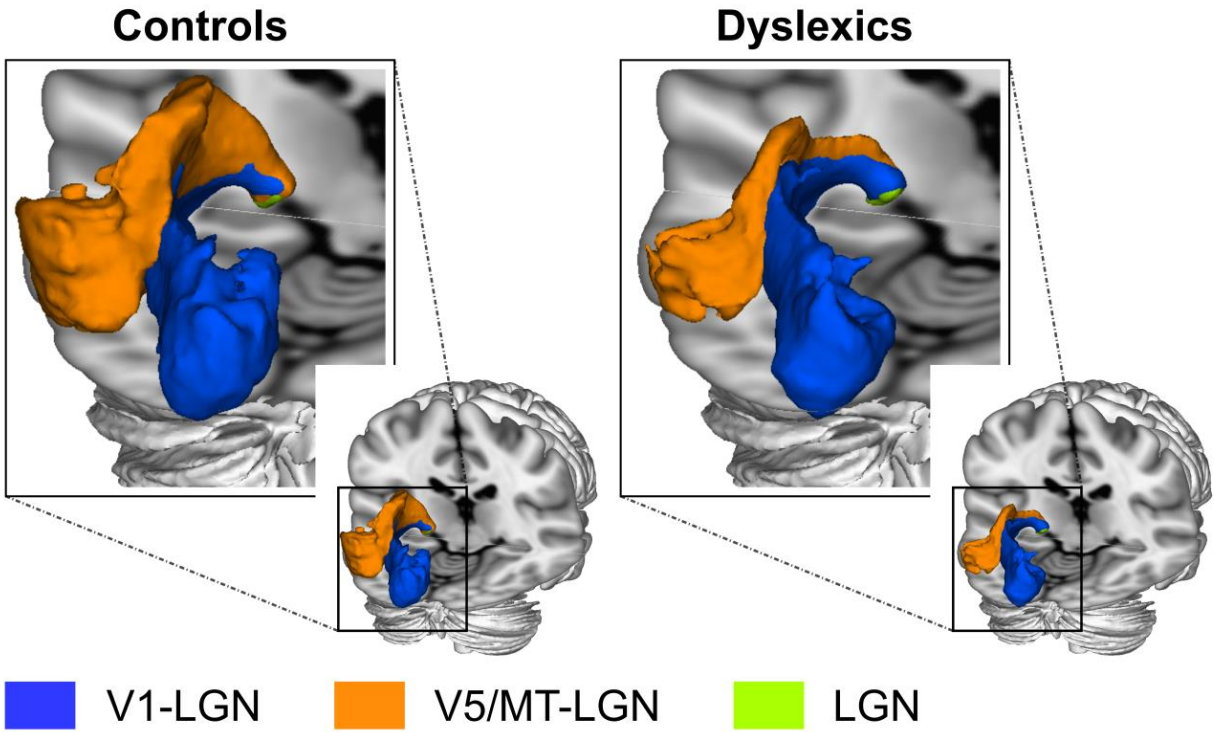

**Figure 3.** Three-dimensional representation of the left V1-LGN and V5/MT-LGN group averaged tracts in controls (n=12) and dyslexics (n=12) in MNI standard space. The group averaged tracts were thresholded to the same values as in Figure 2.

To formally test whether dyslexics have reduced left-hemispheric V1-LGN and V5/MT-LGN connections, we quantified the connection strength between the left LGN and visual cortical areas left V1 and V5/MT for each individual participant by means of a connectivity index. The connectivity index was defined as the log-normalized streamline count in the LGN target mask resulting from probabilistic tractography from cortical seed areas V1 and V5/MT. A 2×2 mixed-model ANOVA on LGN connectivity indices with group (controls vs. dyslexics) as between-subjects factor and cortical seed area (V1 vs. V5/MT) as within-subjects factor (Figure 4A) showed a significant main effect of cortical seed area [$F(1, 22) = 41.85$, $P < .001$, $\eta^2 = .61$], with higher connectivity indices for the connection left V1-LGN than the connection left V5/MT-LGN. This is in accordance with the fact that a vast majority of LGN neurons are connected with V1 [reviewed in refs. 19, 20]. In line with our hypothesis, we found a significant main effect of group, indicating a general reduction in LGN connections in dyslexics as compared to controls [$F(1, 22) =$

Altered Structural Connectivity of the Left Visual Thalamus in Developmental Dyslexia

4.28, $P = .05$, $\eta^2 = .16$]. In addition, the analysis revealed a significant interaction between group and cortical seed area [$F(1, 22) = 4.97$, $P = .036$, $\eta^2 = .07$]. Planned comparisons (one-tailed independent t-tests; Bonferroni corrected) showed that individuals with dyslexia had significantly lower connectivity indices for the connection left V5/MT-LGN as compared to controls [$t(22) = 3.13$, $P = .005$, $d_s = 1.28$] (Figure 4A). Conversely, there was no difference in the connectivity indices for the left V1-LGN connection between controls and dyslexics [$t(22) = .24$, $P = .82$, $d_s = .10$].

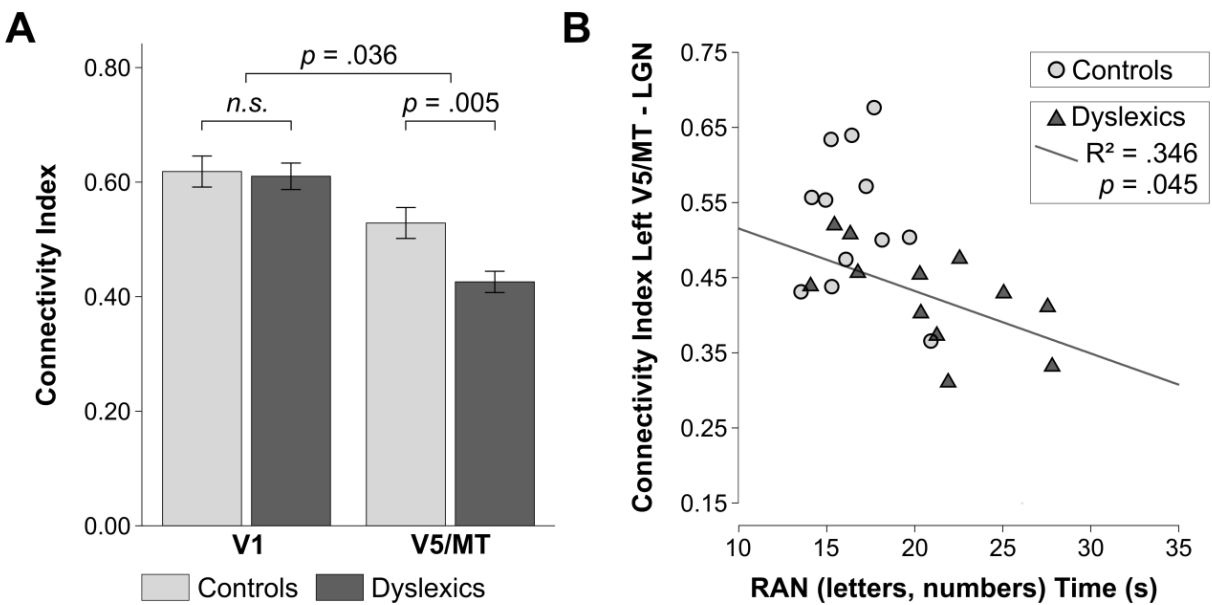

**Figure 4.** Structural connectivity of the LGN in the left hemisphere in controls (n=12) and dyslexics (n=12), and its behavioral relevance for a key dyslexia symptom. **A.** LGN connectivity indices in the left hemisphere in controls and dyslexics obtained from probabilistic tractography using a volume-based atlas for defining cortical seed areas V1 and V5/MT. Error bars represent ± 1 SEM. **B.** The strength of V5/MT-LGN connectivity correlated negatively with the time needed to rapidly name letters and numbers in dyslexic but not in control participants. Rapid naming abilities were measured with the standard diagnostic test for RAN [27].

The difference between dyslexic and control participants in the strength of left V5/MT-LGN connections cannot be explained by differences in the size of seed or target masks between groups. Firstly, we corrected the streamline count in the LGN target masks for the volume of the respective seed mask. Secondly, and even more importantly, there was no significant difference in the mean volume of the LGN target masks [$t(22) = .98$, $P = .34$] nor the V5/MT seed masks [$t(22) = -.11$, $P = .92$] between groups (Table S2).

The finding of reduced structural connections in the direct left V5/MT-LGN pathway in dyslexics adds two fundamental novel contributions to the field. First, while Livingstone et al. [10] have shown histological alterations at the level of the LGN in several post-mortem cases with dyslexia, we here showed that the connections between the LGN and the cerebral cortex were reduced in dyslexics in-vivo. This finding is particularly interesting, because it parallels findings in animal models, where the induction of cortical microgyria, which are similar to those discovered in post-mortem brains of dyslexics, led to a severe reduction in thalamo-cortical and cortico-thalamic connections [13, reviewed in ref. 7]. Second, our study revealed a specific reduction in left-hemispheric cortico-subcortical connections between visual area V5/MT and the LGN in dyslexics, while the connections between V1 and the LGN were spared. Such a specific reduction is informative about the possible functional roles of structural alterations in the early visual pathway for dyslexia symptoms.

A prominent model of dyslexia proposes that subcortical sensory alterations as found in post-mortem studies and animal models are not related to core traits of dyslexia, such as poor reading and slow naming of letters and numbers [7]. Instead, subcortical sensory alterations are thought to solely explain sensory and motor symptoms that are only occasionally associated with developmental dyslexia. Contrary to this assumption, it has recently been proposed that slow naming and poor reading comprehension in developmental dyslexia relate to sensory thalamus dysfunction [16]. To test these opposing hypotheses, we correlated the connectivity indices of the tract for which we had found alterations in the dyslexics (i.e. V5/MT-LGN) with the composite scores on rapid automatized naming (RAN) [27] for letters and numbers as well as the reading comprehension scores. We computed one-tailed Pearson's correlations with Bonferroni correction for two tests (i.e., the correlation of left V5/MT-LGN connectivity once with rapid naming abilities and once with reading comprehension). Our correlation analyses revealed, in dyslexic participants only, a significant negative correlation between the strength of left V5/MT-LGN connections and the time needed to name letters and numbers aloud [$R = -.588$, $P = .045$] (Figure 4B). In control participants, there was no correlation with rapid naming abilities [$R = -.202$, $P = .530$] (Figure 4B). The significant correlation in dyslexics indicated that dyslexic participants with weaker V5/MT-LGN

Altered Structural Connectivity of the Left Visual Thalamus in Developmental Dyslexia

connections had more severe rapid naming deficits. There was no significant correlation between the strength of left V5/MT-LGN connections and participants' reading comprehension scores, neither in dyslexic [*R* = -.213, *P* = .507] nor in control participants [*R* = .076, *P* = .814]. The behavioral correlation between the strength of left-hemispheric V5/MT-LGN connections and a dyslexia diagnostic score (i.e. rapid naming ability) implies a behavioral relevance for V5/MT-LGN alterations for a core dyslexia symptom. Rapid naming performance is one of the strongest predictors of reading ability, and deficits in rapid naming ability present a core symptom of dyslexia in childhood throughout adolescence and in adulthood [28-30, reviewed in ref. 15].

One feature of studies with modest sample sizes is that the results are prone to variation based on minute analytical changes [31]. In a next step, we therefore aimed to reproduce the results of our tractography analysis using a surface-based approach [32] to define visual cortical areas V1 and V5/MT (see Methods). While volume-based atlases are widely used in neuroimaging research, surface-based atlases are thought to yield a higher anatomical mapping accuracy [32, 33]. We found qualitatively the same results as with the volume-based atlases, i.e. a significant reduction of left-hemispheric V5/MT-LGN connections in dyslexics and a significant correlation of the connectivity strength of this pathway with rapid naming abilities in dyslexics only (see Figure S1).

Area V5/MT is critical for the perception of visual motion [reviewed in ref. 34]. Multiple studies have found aberrant motion perception in individuals with dyslexia and pre-readers at familial risk for the disorder [35-37]. In addition, targeted motion perception training improves reading ability in children and adults with dyslexia [36]. However, there are also indications that area V5/MT plays a role for rapid naming abilities: A recent transcranial direct current stimulation (tDCS) study in adult dyslexics showed that anodal stimulation of left area V5/MT (i.e., facilitating cortical V5/MT activity) resulted in a significant improvement in dyslexics' rapid naming abilities for numbers and a trend towards improvement for letters [38]. In addition, a structural MRI study in pre-readers at familial risk for dyslexia found that the gray matter volume of a left-hemispheric occipito-temporal region, which coincides with the anatomical location of area V5/MT, correlated with children's rapid naming abilities [39]. Our finding of a behavioral correlation between the strength of left-hemispheric V5/MT-LGN connections and rapid naming abilities in dyslexics suggests that dyslexia symptoms might not only be linked to area V5/MT or the visual dorsal stream but might be associated with alterations already present at the level of direct V5/MT-LGN connections.

Altered Structural Connectivity of the Left Visual Thalamus in Developmental Dyslexia

The observed correlation between left V5/MT-LGN connectivity strength and rapid naming abilities in dyslexics was not predicted by purely cortical models of dyslexia [see e.g. 7]. The result, however, converges well with recent findings in the auditory modality [16, 40-42]. Most important in the context of the present study, an fMRI study showed lower responses of the left auditory sensory thalamus (i.e., the medial geniculate body, MGB) in dyslexics as compared to controls for recognizing phonemes in contrast to other speech features [16]. Crucially, the amount of this task-dependent left MGB modulation predicted rapid naming abilities for letters and numbers in dyslexics. This previous fMRI and the present study together suggest that rapid naming difficulties in dyslexia are associated with aberrant function of the sensory thalamus and its connections – both in the visual and the auditory modality. The findings also imply that dyslexia might be best explained by a combination of cortical and subcortical sensory accounts of developmental dyslexia [e.g., 16, 40-43] into a comprehensive cortico-subcortical framework.

An interesting question in light of the current findings regards the morphological origin of the reduced left-hemispheric V5/MT-LGN connections in dyslexics. Studies in non-human animals have shown that the direct (V1-bypassing) geniculate projections to area V5/MT originate predominantly in the koniocellular layers of the LGN [21, 23], with only occasional V5/MT relay cells also observed in parvocellular and magnocellular layers. In contrast, histological alterations in post-mortem brains of dyslexics have been observed specifically in the magnocellular layers of the LGN [10]. This apparent discrepancy might at least partly be explained by the fact that most of our current knowledge of the different geniculocortical pathways stems from studies in non-human mammals [19]. As perfect homology between species cannot be assumed [19], it is theoretically possible that the direct LGN-V5/MT connection in humans also comprises strong magnocellular components. In addition, there is currently not much knowledge about the potential feedback connections from area V5/MT to the LGN in any species. There is evidence that thalamic response properties are heavily influenced by cortico-thalamic feedback [reviewed in ref. 44], and that such top-down modulation of early sensory structures is dysfunctional in individuals with dyslexia [16, 41]. In non-human primates, cortico-thalamic feedback from area V5/MT modulates early visual processing in magnocellular, parvocellular and koniocellular LGN cells [45, reviewed in ref. 46]. Thus, our findings could potentially reflect a reduction in feedback connections from left area V5/MT to the magnocellular layers of the LGN. However, at present this is speculative due to the rather low spatial resolution of diffusion-weighted MRI data and because probabilistic tractography does not give information about the direction of resolved connections.

Although the findings give first insight in the potential functional role of subcortical sensory alterations in the visual pathway of dyslexics, the exact nature of how these alterations lead to dyslexia symptoms remains speculative. We favor two potential explanations. First, successful reading and rapid naming involve rapid attentional shifts towards successive visual stimuli [e.g., 47, 48]. Such attentional shifts are known to be controlled by fronto-parietal areas of the dorsal processing stream [49], for which area V5/MT serves as one of the main input structures [reviewed in ref. 34]. Thus, reduced V5/MT-LGN connectivity in dyslexics could result in deficient attention mechanisms through inefficient interactions with these attention-related dorsal stream areas. Second, we have previously hypothesized that dyslexia might be characterized by inefficient top-down modulation of the sensory thalamus to fast-varying predictable stimuli such as speech [16]. This hypothesis was based on findings in the auditory modality [15], but we speculate that similar processes might occur in the visual modality for the fast-varying predictable articulatory movements associated with speech. A potential dysfunction of such top-down modulation of the visual sensory thalamus could explain deficits in speechreading in dyslexia [e.g., 50, 51]. The RAN is a multi-componential measure, which includes access and retrieval of phonological features [30]. Although phonology is often seen as a purely auditory process, it is likely that the brain uses any possible cues to represent speech in the brain including the always associated articulatory movements [e.g., 52, 53]. Thus, we speculate that the RAN scores might be related to the reduced V5-LGN connectivity in dyslexics through a deficit in accessing visual articulatory speech features [16].

Developmental dyslexia has a higher prevalence in males than in females with a 3:1 ratio [3]. In the present study, all of the recruited participants were male. We therefore cannot make claims about female dyslexics as previous studies have shown gender-specific differences in dyslexia [e.g., 54]. In our study we focused on the left hemisphere given evidence that dyslexia is associated with functional and structural alterations particularly of left-hemispheric regions [16, 17, reviewed in ref. 3]. An exploratory analysis of right-hemispheric LGN connectivity (Supplementary Figure S2) revealed no significant reduction in V5/MT-LGN connectivity in dyslexics as compared to controls, suggesting a hemispheric specificity. However, this result has to be taken with extreme caution because the right-hemispheric seed regions were of different sizes in the two groups

Altered Structural Connectivity of the Left Visual Thalamus in Developmental Dyslexia

(Supplementary Table S2); and the interaction between group and cortical seed area in the surface-based analysis (but not the volume-based analysis) was marginally significant ($P$ = .071). Future studies with a higher sample size will likely help to reveal potential hemispheric differences in thalamo-cortical connectivity in dyslexia.

## Conclusions

This study is the first to show structural alterations in visual subcortical sensory pathways in developmental dyslexia. Additionally, it gives important insight into the functional relevance of cortico-thalamic connections for developmental dyslexia, both because of the specificity of the reduction for the LGN-V5/MT connection as well as the behavioral correlation with a key symptom of dyslexia. Together with the few previous studies on subcortical sensory structures and function in dyslexia [10, 11, 16, 40-42], the results imply that the currently predominant approach to investigate dyslexia brain mechanisms predominantly at the cerebral cortex level might not be sufficient for a full understanding of key symptoms of the disorder [e.g., 6, reviewed in refs. 8, 9]. Our study emphasizes the need and paves the way for unraveling the contributions of subcortical sensory pathways to dyslexia with sophisticated neuroimaging approaches with high spatial resolution. We expect that in the future such approaches will lead to a better understanding of dyslexia symptoms within a comprehensive cortico-subcortical framework.

**Authors Contributions.** C.M.A. and K.v.K. designed the experiment. C.M.A and A.A. performed the experiment and analyzed the data. C.M.A. and K.v.K. wrote the manuscript.

**Acknowledgments.** We are grateful to our participants for their time and effort to participate in this study. We thank Morgan Blumberg for her contribution to the LGN segmentations. We thank Begoña Díaz, Domenica Wilfling, Elisabeth Wladimirow and Florian Hintz for providing the structural MRI data and the dyslexia diagnostic scores. Finally, we would like to thank three anonymous reviewers as well as Stefan Kiebel and Louise Kauffmann for their valuable comments on an earlier version of this manuscript. The work was supported by a Max Planck Research Group grant and an ERC-Consolidator Grant (SENSOCOM, 647051) to K.v.K. The authors declare no competing financial interests.

Altered Structural Connectivity of the Left Visual Thalamus in Developmental Dyslexia

Altered Structural Connectivity of the Left Visual Thalamus in Developmental Dyslexia

Altered Structural Connectivity of the Left Visual Thalamus in Developmental Dyslexia

## STAR Methods

**Contact for Reagent and Resource Sharing**

Further information and requests for resources and reagents may be directed to, and will be fulfilled by the Lead Contact, Christa Müller-Axt (muelleraxt@cbs.mpg.de).

**Experimental Model and Subject Details**

Participants

Twenty-four adult German speakers were recruited for the current study. The sample included 12 participants with developmental dyslexia and 12 control participants (see Table S1 for participants' demographic data). Both groups were matched in chronological age, sex, educational level, handedness and non-verbal IQ. In the dyslexia group, 6 participants had a prior diagnosis of developmental dyslexia, while the other 6 participants reported having persistent reading and spelling problems since childhood. Group assignments were confirmed by tests on reading speed and comprehension [56] and spelling [57]. In addition, skills of rapid automatized naming (i.e., RAN for numbers and letters) [27] were assessed. Participants with dyslexia scored lower than controls on the literacy tests as well as on rapid automatized naming for letters and numbers. The scores on the diagnostic tests of dyslexia are summarized in Table S1. Written informed consent was obtained from all participants. The study was approved by the ethics committee of the Medical Faculty, University of Leipzig, Germany.

**Method Details**

Participant Inclusion Criteria

Participants had to meet the following inclusion criteria: (i) no prior history of neurological and psychiatric disorders, (ii) free of psychostimulant medication, (iii) no coexisting neurodevelopmental disorders other than dyslexia (such as dyscalculia, dyspraxia), (iv) no hearing disabilities, and (v) a non-verbal IQ within the normal range (≥ 85). The first four criteria were assessed via participants' self-reports. Non-verbal IQ was assessed with the Raven's advanced progressive matrices test [58].

All participants included in the present study (N=24) were native German speakers and were part of a larger sample (N=28) in a previous fMRI study [16]. Of the 14 participants in each group in the fMRI study, two dyslexics and one control subject were excluded from the present study because no diffusion MRI data could be obtained. An additional control subject was excluded because the non-verbal IQ was below the normal range (i.e., < 85).

Acquisition of High Resolution 7T MRI Data

Ultra-high resolution whole-brain anatomical images were acquired on a 7 Tesla Magnetom MRI system (Siemens Healthineers, Erlangen, Germany) using a 24-channel head array coil (NOVA Medical Inc., Wilmington MA, USA). We employed a 3D MP2RAGE sequence [59] with the following imaging parameters: 700 μm isotropic resolution, TE = 3.04 ms, TR = 8250 ms, $TI_1$ = 1000 ms, $TI_2$ = 3300 ms, flip angle$_1$ = 7°, flip angle$_2$ = 5°, FoV = 224 × 224 × 168 mm$^3$, GRAPPA [60] acceleration factor = 2. The MP2RAGE sequence included two read-outs at different inversion times, from which a quantitative map of $T_1$ relaxation times per voxel was calculated. These maps are well suited for brain segmentations as they provide excellent tissue contrast between white and gray matter [59]. The acquisition took approximately 13 minutes.

Acquisition of Diffusion MRI Data

Diffusion-weighted MRI (dMRI) data were acquired on a 3 Tesla Tim Trio MRI system using a 32-channel head coil (Siemens Healthineers, Erlangen, Germany). We employed a twice-refocused spin-echo echo-planar imaging (EPI) sequence [61] with the following imaging parameters: voxel size = 1.72 x 1.72 x 1.7 mm$^3$, TE = 100 ms, TR = 12.9 s, FoV = 220 x 220 mm$^2$, 88 axial slices covering the whole brain, no inter-slice gap. Diffusion-weighted MRI data were acquired for 60 diffusion-encoding gradient directions with a b-value of 1000 s/mm$^2$. In addition, seven interspersed anatomical reference images without diffusion-weighting (b-value = 0 s/mm$^2$) were obtained for off-line motion correction. The dMRI acquisition was accelerated using partial Fourier imaging (factor 6/8) and parallel imaging [GRAPPA; 60] with an acceleration factor of 2. Fat saturation was applied using a spectral saturation pulse. The dMRI sequence took approximately 16 min. In addition, a $T_1$-weighted structural 3D image was acquired as anatomical reference on the same MRI system (MPRAGE, TE = 3.46 ms, TR = 1300 ms, TI = 650 ms, flip angle = 10°, 1 mm isotropic resolution, two averages).

Preprocessing of Diffusion MRI Data

For preprocessing, the 3 Tesla $T_1$-weighted structural images were skull-stripped and rigidly aligned with the Talairach orientation [62] using the software package LIPSIA (http://www.cbs.mpg.de/institute/software/lipsia). To estimate motion correction

Altered Structural Connectivity of the Left Visual Thalamus in Developmental Dyslexia

parameters for the dMRI data, we used the seven reference images without diffusion-weighting and rigid-body registration [63], implemented in FSL (version 5.0, FMRIB Software Library, University of Oxford http://www.fmrib.ox.ac.uk/fsl). Motion correction parameters were interpolated for all 67 volumes and combined with a global registration to the $T_1$ anatomy (in Talairach orientation) using rigid-body registration. The estimated motion correction parameters were then used to correct the gradient directions of each dMRI volume. The registered dMRI volumes were resampled to an isotropic voxel resolution of 1.72 mm and the background was masked with the skull-stripped $T_1$-image. Finally, a diffusion tensor was fit to each voxel and fractional anisotropy (FA) maps were computed.

Segmentation of the Lateral Geniculate Nucleus
Individual ultra-high resolution 7 Tesla MRI quantitative $T_1$ maps were used to manually segment the left LGN in each participant (Figure 1A). In coronal view, the LGN is ventrally adjoined by the hippocampal sulcus (cerebral spinal fluid, CSF). Dorsolaterally, the LGN is surrounded by the white matter fibers that form the triangular area (zone of Wernicke) [64]. Given the lower $T_1$ relaxation times of the surrounding white matter and the high $T_1$ relaxation times of CSF [65], the LGN can be clearly distinguished on quantitative $T_1$ maps. Manual segmentations were performed in FSLView (http://fsl.fmrib.ox.ac.uk/fsl/fslview) by two independent raters who were both blind to participants' group assignment.

*Standardized Segmentation Procedure*
In order to standardize the LGN segmentation procedure between raters and across participants, we first computed the histogram of $T_1$ relaxation times for each participant (number of bins = 1000, bin size = 4). This yielded clear peaks of $T_1$ relaxation times in grey and white matter for each participant. We then loaded the MRI volume of each participant in the viewer and set the minimum intensity to half a standard deviation below the individual white matter relaxation time peak, while the maximum intensity was set to half a standard deviation above the individual grey matter $T_1$ relaxation time peak (corresponding to 88% of the peak $T_1$ relaxation time intensities assuming a Gaussian normal distribution). This was done to optimize the visibility of the LGN. Segmentations were then performed in coronal view aided by the sagittal and transverse views. All segmentations were performed in randomized order across participants. Finally, the LGN masks of both raters were conjoined, such that the final

LGN masks only comprised those voxels that were segmented by both raters (Figure 1A).

*Inter-rater Reliability*

Inter-rater reliability for the manual LGN segmentations was assessed by computing the dice coefficient as twice the amount of shared voxels between both raters' LGN masks divided by the total number of voxels in both masks: (2 × mask_1 ∩ mask_2) / (mask_1 + mask_2), wherein mask_1 and mask_2 refer to the LGN masks of rater 1 and rater 2, respectively [66]. The obtained coefficient yields a measure of the amount of agreement between the two raters and ranges from 0 (no agreement) to 1 (perfect agreement). The agreement between raters was high – both for the LGN segmentations in control participants (0.86 ± 0.04, mean ± SD) as well as in dyslexic participants (0.85 ± 0.04).

*Registrations to Individual dMRI Data*

Conjoined LGN masks were registered to the individual dMRI data using a combination of linear and non-linear registrations computed with FSL. To facilitate registration accuracy, we used the individual co-registered skull-stripped uniform images of the MP2RAGE sequence rather than the quantitative $T_1$ maps as input for these registrations. The $T_1$-weighted structural images (aligned with the dMRI data) served as registration target. The LGN masks were then warped into the individual space of the dMRI data by applying the obtained linear and non-linear registration parameters. All registrations were visually inspected for misalignments. Finally, the registered individual LGN masks were thresholded at an intensity of 0.4 (after linear interpolation to the target image) to preserve the volumes of the conjoined LGN masks. The group-specific LGN mask volumes after registration and thresholding are summarized in Table S2.

*Registrations to MNI Standard Space*

For visualization purposes, we normalized the untresholded conjoined LGN masks in dMRI data space to MNI standard space (using the MNI 1 mm brain template as reference). We then averaged the LGN masks within each group to derive a voxel-wise probability map of anatomical overlap across participants (Figure 1B).

Altered Structural Connectivity of the Left Visual Thalamus in Developmental Dyslexia

Definition of Cortical Regions of Interest

We employed both a volume-based and a surface-based approach to define the cortical regions of interest (ROI) left V1 and V5/MT. Both approaches are described in detail below.

*Volume-based Definition of Left V1 and V5/MT*

We derived volume-based probabilistic atlases of left V1 and V5/MT from the Juelich Histological Atlas [67], as implemented in FSL. Both probabilistic atlases were in MNI standard space with a voxel size of 1 mm$^3$.

The unthresholded probabilistic atlases of area V1 [68] and V5/MT [69, 70] cover large spatial extents due to the inter-individual anatomical variability of these areas. We therefore used the following procedure to find appropriate thresholds for the probabilistic atlases to confine the final left V1 and left V5/MT ROIs to anatomically plausible volumes: we first thresholded the probabilistic atlases at different percentages of overlap in steps of 5% from 5% to 95%. We then transformed all of these differently thresholded V1 and V5/MT atlases to each participant's dMRI data. We therefore registered the MNI 1 mm brain template to the individual $T_1$-weighted structural images, which were aligned with the diffusion-weighted images. Linear and non-linear registration was performed with FSL using default parameters. The probabilistic atlases were then warped to the individual dMRI data by applying the obtained linear and non-linear registration parameters. Next, the gray-matter areas of the registered V1 and V5/MT maps were computed using voxels inside the brain with FA < 0.2 to guarantee a good quality mask for tractography. Finally, the remaining mean gray matter volumes of the atlases across all participants for each initial threshold were compared to anatomical volume estimates of left V1 and left V5/MT that have been reported in the literature, and the threshold that most closely corresponded to the reported volume estimate of the respective brain area was selected.

Based on post-mortem measurements, Andrews et al. [18] reported volumes of left V1 = 3185 - 7568 mm$^3$. We focused on the high end of the reported V1 volume range to assure that all voxels that are part of area V1 were included in the individual V1 ROIs and to partially account for volumetric tissue shrinkage in post-mortem preparations [71]. We therefore chose a threshold of 60 % for the probabilistic atlas of left V1, which resulted in a mean volume of left V1 = 6532 ± 792 mm$^3$ across participants in dMRI data space. The group-specific left V1 ROI volumes are summarized in Table S2 (volume-based cortical ROIs).

Using a functional localizer, Bridge et al. [24] reported a volume of 2500 mm$^3$ for area V5/MT. We therefore chose a threshold of 10 % for the probabilistic atlas of left V5/MT, which resulted in a mean volume of left V5/MT = 2765 ± 495 mm$^3$ across participants in dMRI data space. The group-specific volume-based left V5/MT ROI volumes are summarized in Table S2 (volume-based cortical ROIs).

The comparably great difference in chosen threshold for left V5/MT (10%) and left V1 (60%) atlases can be explained by a greater inter-individual structural variability in V5/MT location compared to V1 location. The maximum anatomical overlap in the probabilistic atlas of left V5/MT is 54% as compared to 100% for left V1.

*Surface-based Definition of Left V1 and V5/MT*

Surface-based atlases of left V1 and V5/MT were derived from a recently published and cross-validated atlas based on fMRI [32]. We used the provided maximum probability maps (i.e., indicating the most probable region for any given point) instead of the full probability maps (i.e., indicating the likelihood that a given point is part of any region) in order to avoid probability and thus volume thresholding. As the surface-based atlas features separate maximum probability maps of both ventral and dorsal left V1, these two maps were conjoined to obtain a surface-based map covering entire left V1.

In order to map the surface-based atlases of V1 and V5/MT on each individual subject, we first reconstructed each participant's cortical surface using the software package FreeSurfer (https://surfer.nmr.mgh.harvard.edu/). Each participant's skull-stripped T$_1$ image (aligned with the dMRI data) served as input for the reconstruction process. Subsequently, cortical surfaces were imported into the software AFNI (https://afni.nimh.nih.gov/) and resampled to match the template brain surface of the atlas using @SUMA_Make_Spec_FS. Left V1 and V5/MT maximum probability labels were then mapped onto the individual surfaces and vertex coordinates of the labeled surface points were converted into voxel coordinates and marked in the individual brain volume. The surface-based masks were then dilated by one voxel (1 mm) in each dimension. This was done to facilitate tractography by ensuring closer proximity of the surface-based V1 and V5/MT ROIs to the surrounding white matter. Finally, the surface-based V1 and V5/MT ROIs were resampled to match the resolution of the dMRI data. This procedure resulted in a mean volume of left V1 = 4385 ± 535 mm$^3$ and left V5/MT = 958 ± 204 mm$^3$ across participants in dMRI data space. The group-specific surface-based left V1 and left V5/MT ROI volumes are summarized in Table S2 (surface-based cortical ROIs).

Altered Structural Connectivity of the Left Visual Thalamus in Developmental Dyslexia

Probabilistic Tractography

We computed voxel-wise estimates of the fiber orientation distribution [72] from the preprocessed dMRI data using FSL. We estimated the distribution of up to two fiber orientations for each voxel, given the b-value and resolution of the dMRI data [73]. Probabilistic tractography was performed in individual dMRI data space using FSL with default parameters. This produces an estimate of the probability and strength of the most likely location of a pathway [73]. We used modified Euler streamlining with the visual cortical areas (i.e., left V1 or V5/MT) as seeds and the participant-specific left LGN as both waypoint and termination mask to compute the connectivity between the LGN and the respective cortical area. All analyses were done separately for each pair of seed and target region. Tractography was only computed from the cortical region to the LGN to better detect possible non-dominant connections to the cortex, which might be missed by the algorithm when seeding in the LGN. In branching situations, probabilistic tractography has the tendency to miss the non-dominant connection (false negative results), which can be reduced by seeding in cortical areas [74].

*Connectivity Index*

For each participant and pair of seed and target region, we computed a connectivity index, $I$, which was determined from the number of sample streamlines from each seed that reached the target [75]. As this number strongly depends on the number of voxels in the respective seed mask, we normalized the connectivity index, $I$, according to the following equation:

$$I = \frac{log(waytotal)}{log(5000 \times V_{seed})} \,,$$

wherein *waytotal* refers to the number of streamlines from a given seed that reached the target (i.e., numeric output of the tractography algorithm), *5000* refers to the number of generated sample streamlines in each seed voxel, and $V_{seed}$ denotes the number of voxels in the respective seed mask. As the connectivity indices cannot be expected to be normally distributed across participants, we computed the logarithmic scaling (log) of each term of the equation to transform the connectivity index into a normally distributed variable (ranging between 0 and 1).

2626

Altered Structural Connectivity of the Left Visual Thalamus in Developmental Dyslexia

*Tracking Consistency*

To evaluate whether connections were consistently resolved in each participant, they had to meet three criteria: first, we evaluated the spatial consistency of the resolved connections across participants by visual inspection. This was done to assure that the reconstructed pathways followed the known anatomical literature priors [see e.g. 25]. Second, for the binary decision whether a specific connection was strong enough to be reliably detected by tractography, we regarded a connection between two brain areas as detected if at least 10 of the generated sample streamlines in a given seed region reached the target [76]. Finally, we computed the mean and standard deviation (SD) of the connectivity indices (i.e., log-normalized streamline counts) for both the left-hemispheric V1-LGN and V5/MT-LGN connection separately for each group. A connection of a participant was considered inconsistent if his or her connectivity index was > 2.5 SDs away from the group mean connectivity index for the respective connection. All three criteria were fulfilled for each participant and each connection considered in this study.

*Group Averaged Tracts in MNI Standard Space*

The output images of the probabilistic tractography for each connection (visitation maps) were normalized by logarithmic transformation (of all values > 0) and division by the logarithm (log) of the total number of generated streamlines in each seed mask. This was the same normalization procedure as described in detail in the Methods for the connectivity indices. The resulting images contained the log-normalized number of streamlines per voxel, scaled between 0 and 1. The log transformation helped to approach a Gaussian normal distribution of the initially not normally distributed visitation values and is mandatory for the normalization to the MNI 1 mm brain template, which includes linear interpolation. In MNI space, each connection was averaged within controls and dyslexics to derive group averaged tracts for left-hemispheric V1-LGN and V5/MT-LGN connections (Figure 2 and Figure 3).

**Quantification and Statistical Analysis**

Statistical analysis was performed using Matlab (version 8.6, The MathWorks Inc., MA, USA) and IBM SPSS Statistics (version 22, IBM Corporation, NY, USA). For all statistical tests, n was defined as the number of participants in each group (i.e., controls vs. dyslexics) that were included in the respective analysis. Group differences in participants' demographic and diagnostic data as well as ROI volumes were analyzed using independent t-tests. Statistical tests and corresponding parameters including the exact number of n, central tendency (mean) and dispersion (SD) of participants' demographic and diagnostic data as well as ROI volumes are reported in Table S1 and S2, respectively. Group differences in structural LGN connectivity were analyzed using 2x2 mixed-model ANOVAs and independent t-tests, where appropriate. Details on the employed statistical analyses and corresponding parameters are described in the Main Text and in the Figure Legends of Figures 4A, S1A and S2. Data on structural LGN connectivity in Figures 4A, S1A and S2 are expressed as mean ± standard error of the mean (SEM). Behavioral correlations with structural LGN connectivity were calculated using Pearson's correlations. Details on the correlation analyses are described in the Main Text as well as in Figures 4B and S1B and associated Figure Legends. For all statistical tests, the significance level α was defined at 5% (p ≤ .05). Effect sizes for ANOVAs were calculated using eta squared ($\eta^2$) [77]. Effect sizes for independent t-tests were calculated using Cohen's $d_s$ [78]. All measures met the normality assumption as assessed with the Shapiro-Wilk test [79].

Altered Structural Connectivity of the Left Visual Thalamus in Developmental Dyslexia

## Supplemental Data

**Table S1.** Demographic data and diagnostic tests for controls and dyslexics, related to Figure 1, and Figure 4B.

|  | Control group (n = 12) | Dyslexia group (n = 12) | Independent t-tests (df = 22) |
|---|---|---|---|
| Demographic data |  |  |  |
| Mean age ± SD, years | 23.7 ± 2.6 | 24.2 ± 2.4 | NS |
| Male sex | 12 | 12 | – |
| No. right-handed | 11 | 10 | – |
| Education level | 11 undergraduate students, 1 high school diploma | 12 undergraduate students | – |
| Diagnostic tests, mean ± SD |  |  |  |
| Nonverbal intelligence[a] | 110.8 ± 12.8 | 101.0 ± 13.6 | NS |
| Spelling[b] | 102.8 ± 5.6 | 83.1 ± 7.6 | $t = 7.2, P < .001$ |
| Reading speed[c] | 58.3 ± 9.1 | 42.6 ± 6.5 | $t = 4.9, P < .001$ |
| Reading comprehension[c] | 62.9 ± 7.7 | 47.4 ± 4.2 | $t = 6.1, P < .001$ |
| RAN numbers |  |  |  |
|   Time (s) | 16.8 ± 2.4 | 21.2 ± 6.1 | $t = 2.3, P < .05$ |
|   Errors (%) | 0.8 ± 1.3 | 0.2 ± 0.6 | NS |
| RAN letters |  |  |  |
|   Time (s) | 16.4 ± 2.6 | 20.3 ± 3.5 | $t = 3.1, P < .01$ |
|   Errors (%) | 0.3 ± 1.2 | 0.3 ± 0.8 | NS |

RAN = rapid automatized naming
[a]Raven matrices, scores based on standard scores (mean = 100, SD = 15).
[b]Spelling test, scores based on standard scores (mean = 100, SD = 10).
[c]Reading speed and comprehension tests, scores are based on t-standard scores (mean = 50, SD = 10).

Altered Structural Connectivity of the Left Visual Thalamus in Developmental Dyslexia

**Table S2.** ROI volumes for controls and dyslexics in dMRI data space, related to Figure 1A, Figure 2, Figure 4, Figure S1, and Figure S2.

| ROI volumes in mm$^3$, mean ± SD | Control group (n = 12) | Dyslexia group (n = 12) | Independent t-tests (df = 22) |
|---|---|---|---|
| L LGN | 132.4 ± 27.3 | 121.4 ± 27.5 | $t$ = 0.98, $P$ = .34 |
| R LGN | 121.4 ± 23.1 | 113.4 ± 25.4 | $t$ = 0.81, $P$ = .43 |
| Volume-based cortical ROIs | | | |
|   L V1 | 6537.9 ± 915.7 | 6526.9 ± 686.7 | $t$ = 0.03, $P$ = .97 |
|   L V5/MT | 2753.6 ± 436.2 | 2775.6 ± 567.5 | $t$ = 0.11, $P$ = .92 |
|   R V1 | 8018.0 ± 1437.4 | 7107.5 ± 793.9 | $t$ = 1.92, $P$ = .07 |
|   R V5/MT | 2061.0 ± 350.9 | 2364.4 ± 327.1 | $t$ = 2.19, $P$ = .04 |
| Surface-based cortical ROIs | | | |
|   L V1 | 4407.6 ± 545.6 | 4361.5 ± 547.6 | $t$ = 0.21, $P$ = .84 |
|   L V5/MT | 911.0 ± 236.3 | 1004.5 ± 163.5 | $t$ = 1.13, $P$ = .27 |
|   R V1 | 3990.4 ± 470.6 | 4004.3 ± 631.6 | $t$ = 0.06, $P$ = .95 |
|   R V5/MT | 950.3 ± 225.6 | 980.8 ± 191.6 | $t$ = 0.36, $P$ = .72 |

L = left; R = right

Altered Structural Connectivity of the Left Visual Thalamus in Developmental Dyslexia

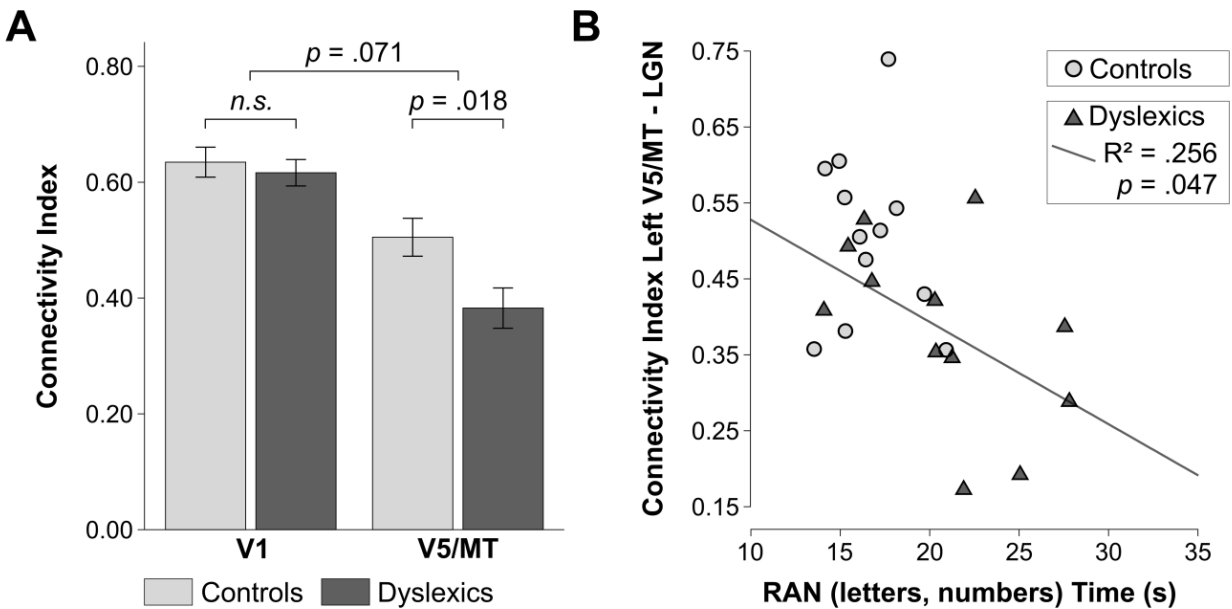

**Figure S1.** Results of the tractography analysis using a surface-based definition of cortical seed areas V1 and V5/MT (see Table S2), related to Figure 4.

Structural connectivity of the LGN in the left hemisphere in controls (n=12) and dyslexics (n=12), and its behavioral relevance for a key dyslexia symptom. **A.** LGN connectivity indices in the left hemisphere in controls and dyslexics obtained from running tractography using a surface-based atlas for defining cortical seed areas V1 and V5/MT. The LGN connectivity indices showed a high agreement with those obtained using the volume-based atlas for tractography: both for the connection left V1-LGN [$R = .94$, $P < .001$; one-tailed Pearson's correlation] and the connection left V5/MT-LGN [$R = .83$, $P < .001$; one-tailed Pearson's correlation]. A 2×2 mixed-model ANOVA on the obtained LGN connectivity indices revealed a significant main effect of cortical seed area [$F(1, 22) = 44.28$, $P < .001$, $\eta^2 = .63$], with higher connectivity indices for the connection left V1-LGN than the connection left V5/MT-LGN. The interaction between group and cortical seed area showed a trend towards significance [$F(1, 22) = 3.6$, $P = .071$, $\eta^2 = .05$]. Planned comparisons (one-tailed independent t-tests; Bonferroni corrected) revealed the same pattern of simple effects as in the analysis with the volume-based atlases: dyslexic participants had significantly lower connectivity indices for the connection left V5/MT-LGN as compared to controls [$t(22) = 2.55$, $P = .018$, $d_s = 1.04$], while there was no difference in the connectivity indices for the connection left V1-LGN [$t(22) = .54$, $P = .60$, $d_s = .22$] between groups. Error bars represent ± 1 SEM. **B.** Consistent with the results of the volume-based analysis (Figure 4B), we found again that the strength of

Altered Structural Connectivity of the Left Visual Thalamus in Developmental Dyslexia

V5/MT-LGN connections correlated negatively with the time needed to rapidly name letters and numbers in dyslexic [$R$ = -.506, $P$ = .047] but not in control participants [$R$ = -.141, $P$ = .331].

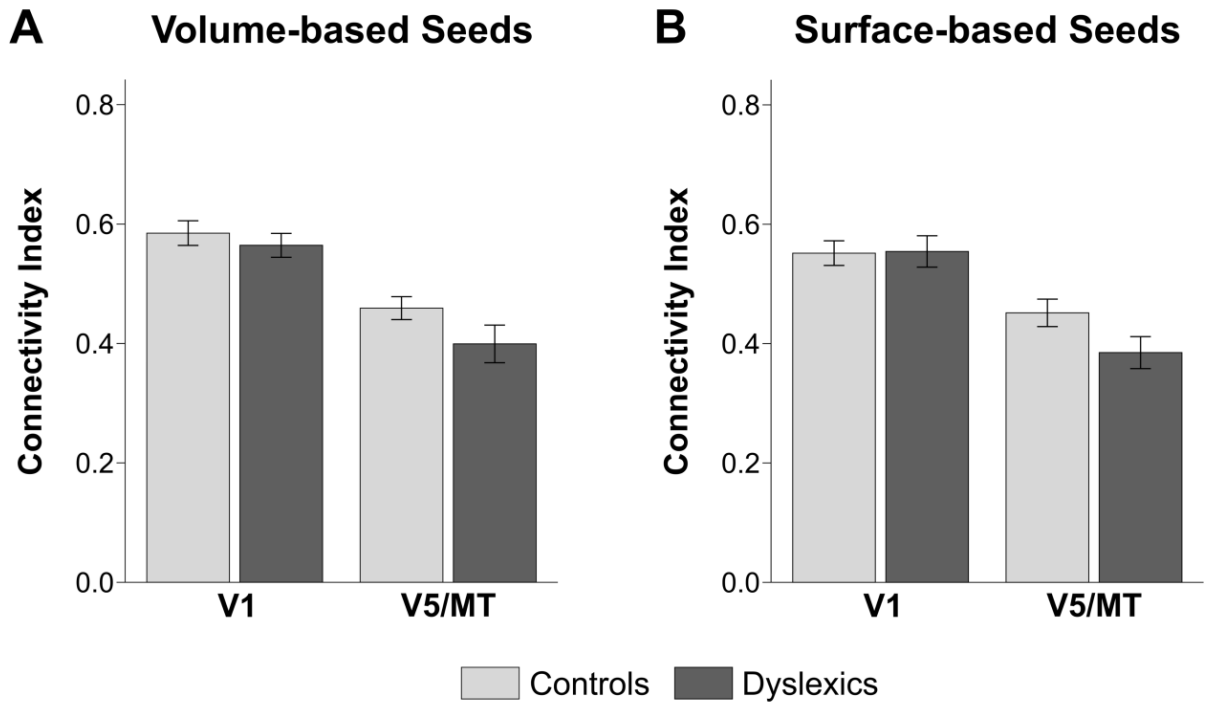

**Figure S2.** Results of the exploratory analysis on right-hemispheric LGN connectivity (see Table S2), related to Figure 4A, and Figure S1A.

Structural connectivity of the LGN in the right hemisphere in controls (n=12) and dyslexics (n=12). We defined right-hemispheric LGN masks by following the same standardized segmentation procedure as in the left hemisphere. Inter-rater reliability for right-hemispheric LGN segmentations was high (mean dice coefficient ± SD; controls: 0.85 ± 0.03, dyslexics: 0.86 ± 0.04). Conjoining both rater's LGN masks resulted in an average right LGN mask volume of 119 ± 22 mm$^3$ in controls and 114 ± 19 mm$^3$ in dyslexics. Conjoined right LGN masks were registered to the dMRI data using the same registrations as for left-hemispheric LGN masks. Volume-based atlases of visual cortical areas right V1 and V5/MT were obtained from the same source as for visual cortical areas left V1 and V5/MT. Thresholds for the unregistered volume-based atlases of right V1 and V5/MT were chosen following the same procedure as for the left-hemispheric volume-based atlases. This resulted in a threshold of 60% for right V1 and 15% for right V5/MT. Probabilistic tractography was performed as described for the left hemisphere.

Altered Structural Connectivity of the Left Visual Thalamus in Developmental Dyslexia

**A.** LGN connectivity indices in the right hemisphere in controls and dyslexics obtained from probabilistic tractography using a volume-based atlas for defining cortical seed areas V1 and V5/MT. The definition of cortical seed areas right V1 and V5/MT resulted in (marginally) significant differences in the amount of grey matter voxels in the seed areas between groups (Table S2). Although we corrected the streamline count in the LGN target masks for the volume of the respective seed mask, significant differences in the size of the seed masks between groups are likely to yield biased estimations of the connectivity indices and the here presented results have to be therefore taken with caution. We calculated a 2 × 2 mixed-model ANOVA on right-hemispheric LGN connectivity indices with group (controls vs. dyslexics) as between-subjects factor and cortical seed area (right V1 vs. right V5/MT) as within-subjects factor. We found a significant main effect of cortical seed area [$F(1, 22) = 58.97$, $P < .001$, $\eta 2 = .72$], with higher connectivity indices for the connection right V1-LGN than the connection right V5/MT-LGN. Both the main effect of group [$F(1, 22) = 2.19$, $P = .15$, $\eta 2 = .09$] and the interaction between group and cortical seed area [$F(1, 22) = 1.08$, $P = .31$, $\eta 2 = .01$] were non-significant. **B.** LGN connectivity indices in the right hemisphere in controls and dyslexics obtained from probabilistic tractography using a surface-based atlas for defining cortical seed areas V1 and V5/MT. Surface-based atlases of visual cortical areas right V1 and V5/MT were obtained from the same source as for visual cortical areas left V1 and V5/MT. A 2 x 2 mixed-model ANOVA on the right-hemispheric LGN connectivity indices yielded a significant main effect of cortical seed area [$F(1, 22) = 54.45$, $P < .001$, $\eta 2 = .68$], with higher connectivity indices for the connection right V1-LGN than the connection right V5/MT-LGN. The interaction between group and cortical seed area showed a trend towards significance [$F(1, 22) = 3.6$, $P = .071$, $\eta 2 = .05$], while the main effect of group was non-significant [$F(1, 22) = 1.20$, $P = .29$, $\eta 2 = .05$]. While we found no significant interaction between group and cortical seed area in the volume-based analysis, the same interaction was marginally significant in the surface-based analysis. As this interaction seemed to be driven by a similar connectivity pattern as in the left hemisphere, we cannot exclude the possibility that a reduction in V5/MT-LGN connectivity in dyslexics might also be present in the right hemisphere. Error bars represent ± 1 SEM.